\documentclass[letter]{aa}  

\newcommand{\filter}[1]{\mbox{\it #1\/}}              
\usepackage{graphicx}
\usepackage{txfonts}
\usepackage{natbib}
\begin{document}

   \title{A common column density threshold for  scattering at 3.6 $\mu$m and  water-ice in molecular clouds}
\titlerunning{Common threshold for 3.6 $\mu$m scattering and water-ice}

   \author{ M. Andersen
          \inst{1,2}
          \and
          W-F. Thi \inst{1}
          \and
          J. Steinacker \inst{1,3}
          \and 
          N. Tothill \inst{4}}
     
            \institute{Univ. Grenoble Alpes, IPAG, F-38000 Grenoble, France
CNRS, IPAG, F-38000 Grenoble, France \and    Aix Marseille Universit\'e, CNRS, LAM (Laboratoire d'Astrophysique de Marseille), UMR 7326, 13388, Marseille, France  \email{morten.andersen@lam.fr}
\and Max-Planck-Institut f\"ur Astronomie, 
K\"onigstuhl 17, D-69117 Heidelberg, Germany
   \and University of Western Sydney, Locked Bag 1797, Penrith 2751 NSW, Australia}


 
  \abstract
   {Observations of scattered light in the 1-5 $\mu$m range have revealed  dust grains in molecular cores with sizes  larger than commonly inferred  for the diffuse interstellar medium. It is currently unclear whether these grains are grown within the molecular cores or are an ubiquitous component of the interstellar medium. }
   {We investigate whether the large grains necessary for efficient scattering at 1-5 $\mu$m are associated with  the abundance of water-ice within molecular clouds and cores.  }
   {We combined water-ice abundance measurements for sight lines through the Lupus IV molecular cloud complex with  measurements of the scattered light at 3.6 $\mu$m for the same sight lines. }
   {We find that there is a similar threshold for the cores  in emission in scattered light at 3.6 $\mu$m  ( $\tau_{9.7}=0.15\pm0.05$, $A_K=0.4\pm0.2$) as  water-ice ($\tau_{9.7}=0.11\pm0.01$, $A_K=0.19\pm0.04$) and that the scattering efficiency increases as the relative water-ice abundance  increases. The ice layer increases the average grain size, which again strongly increases the albedo.}
   {The higher scattering efficiency is partly due to layering of ice on the dust grains. Although the layer can be relatively thin it can enhance the scattering substantially.}


   \maketitle
%

\section{Introduction}
Interstellar dust is a key ingredient of the interstellar medium and strongly affects observations of other objects. 
The determination of  dust properties is therefore fundamental for our understanding of for example the extinction law and the evolution of dust. 
Evidence for grain growth has been suggested both through changes in the extinction law as a function of column density  and from systematic regional differences in the extinction law  \citep[e.g.][]{flaherty,cambresy,ascenso}. 
Furthermore,  grains with a radius of  around 1 $\mu$m  have been  suggested to be present through coreshine ($CS$), scattering in the mid-infrared \citep{steinacker,pagani}, which are larger  than  expected in the diffuse interstellar medium \citep[e.g.][]{MRN}. 
Modelling in one particular case, the molecular core LDN260, showed that the upper limit of the grain size derived by extending a power-law distribution would be no more than 1.5 $\mu$m \citep{andersen}.

Water-ice   is detected for sight lines with an extinction above a threshold of A$_{V}\sim 2-3$ \citep{whittet_01,boogert,whittet_13,chiar_11}, although the threshold appears to be higher in regions with a stronger radiation field \citep{tanaka}. 
Models of grain growth suggest that coagulation is more efficient when the grains are coated with water-ice \citep[e.g.][]{ormel} and  large grains may thus be related to  the abundance of water-ice. 

In this letter we investigate the connection between the large grains observed through scattering at 3.6 $\mu$m and the water-ice abundance in Lupus IV. 
The observed scattered light is a function of  the dust distribution, but also of the geometry between the molecular cloud and the scattering source, the radiation field being scattered, and the strength of the background radiation, all depending on the location in the Galaxy \citep[][Lef\'evre et al. 2014, A\&A in press]{steinacker_new}.
 The large-scale geometric effects, for instance the phase function, and the radiation field, are nearly constant, which reduces the main uncertainties to the relative geometry between the molecular clouds and the radiation field.

In Section 2 we present the compilation of data used for this study, the further selection of suitable sources,  and the measurement of $CS$\footnote{$CS$ refers to the excess scattering over the extinction of the line of sight diffuse background radiation} for each sight line.
Section 3 presents the main results. 
The $CS$ corrected for optical depth effects is derived. The connection between the relative ice  abundance and $CS$ is discussed together with  the impact of ice on the large grains. The relationships between the water-ice, total dust content, and mid-infrared scattering are presented and the implications are discussed. 
Finally, we conclude in Section 4.


\section{Data}

We have compiled data from the literature and archives for the Lupus IV cloud complex. 
\citet{boogert} obtained {\it Spitzer} IRS low-resolution spectra of 25 background stars, complemented with ground-based VLT/ISAAC \filter{K} and \filter{L}- band spectra with a resolving power of R=1500. 
We here used the derived optical depths of the silicate feature at 9.7 $\mu$m, of the water-ice feature at 3.0 $\mu$m, and the extinction measured in the \filter{K} band, adopted from their Tables 2 and 3. The sample probes an extinction range of  0.14 < $A_K$ < 2.46.

\begin{figure}
 \resizebox{0.9\hsize}{!}{\includegraphics{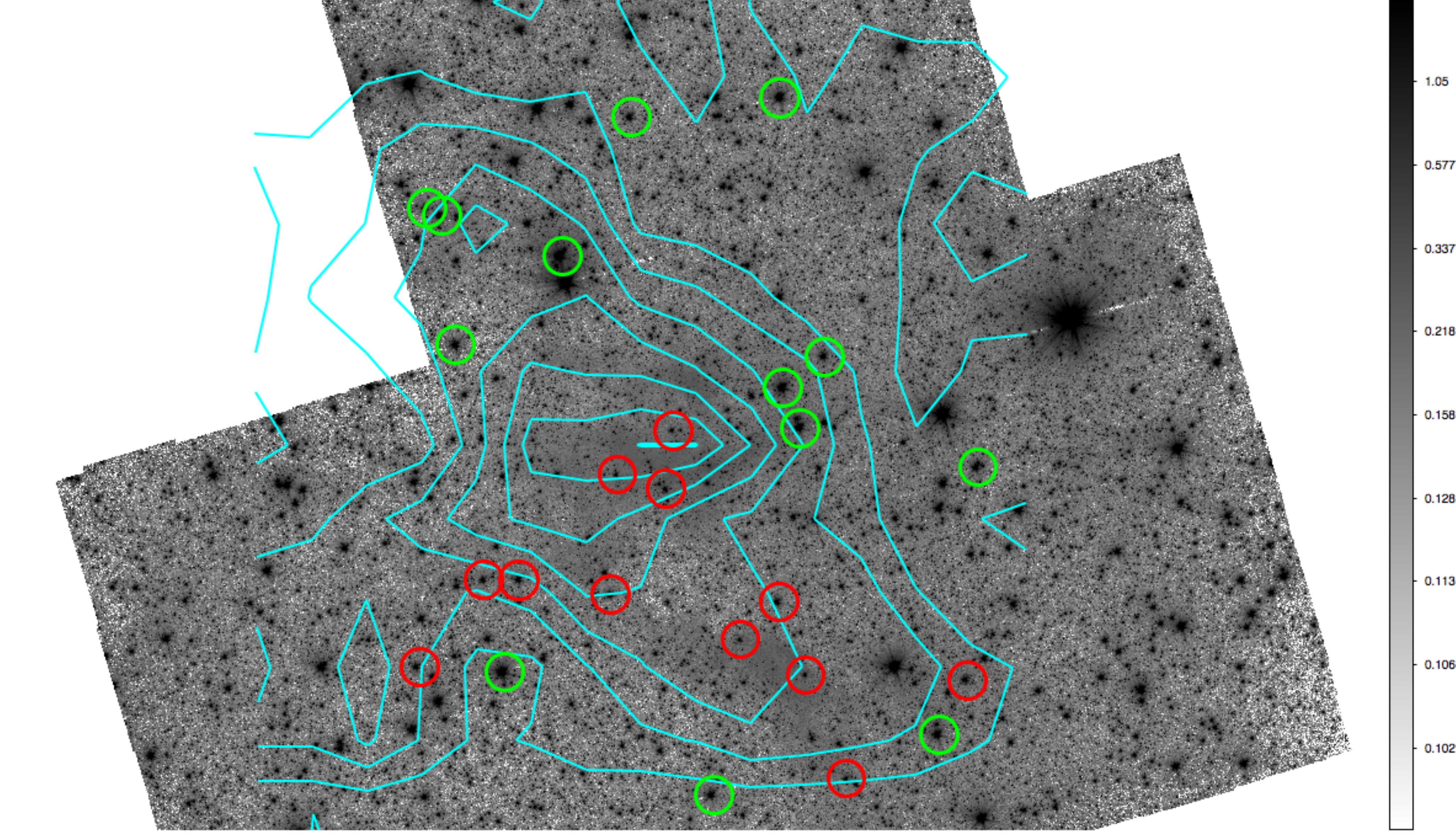}}	
	\caption{{\it Spitzer} 3.6 $\mu$m image of Lupus IV.  The  field of view is 43\arcmin $\times$ 54\arcmin. North is up, east to the left, and scaling is logarithmic. The locations of the background stars for which the optical depth has been measured are  shown as circles  (100 pixels, 60\arcsec) radius,  which is the outer radius of the annulus used to measure $CS$). The red circles show the stars included in the sub-sample used in this study. $^{13}\mathrm{CO}(2-1)$ emission is shown in steps of 1 $\mathrm{K km s^{-1}}$ between 0 and 7 $\mathrm{K\ kms^{-1}}$ as contours.}
	\label{overview}
\end{figure}

A 0.8 square degree area of Lupus IV was observed by {\it Spitzer} through channels 1 and 2 (3.6 $\mu$m and 4.5 $\mu$m, PI Krauss, programme 90071)  of which we used the 3.6 $\mu$m data.
The integration time per frame was 12 seconds with typical coverage of at least eight frames per pixel, providing a shortest integration time per pixel of 96 seconds. 
Finally, $^{13}\mathrm{CO}(2-1)$ data were obtained from \citet{tothill}.  
The data were obtained with the Antarctic Sub-millimeter Telescope and Remote Observatory and the beam size for the  $^{13}\mathrm{CO}(2-1)$ observations was 3.3\arcmin. 
Fig.~\ref{overview} shows the 3.6 $\mu$m observations with the integrated $^{13}\mathrm{CO}(2-1)$ emission shown as contours. The  \citet{boogert} sources are indicated as circles with a radius of 100 pixels, or 60\arcsec.

\subsection{Coreshine surface brightness}
The absorption line studies rely on  a bright background star as the reference source. 
Because of  the  star it is not possible to measure the surface brightness at the exact same position as the extinction measure. 
We  instead measured the surface brightness in an annulus around the star to avoid light from the stellar point-spread-function (PSF). 
The inner and outer radii of the annulus  are 60 and 100 pixels (36-60\arcsec),  corresponding to a physical range of (0.026-0.044 pc). 

The surface brightness within the annulus was determined from a Gaussian fit to the histogram of the pixel value distribution. 
First the stars within the frame were suppressed by PSF subtraction to avoid a bias in the surface brightness distribution. 
Then the Gaussian fit was performed.

The sample  was reduced by excluding  the brightest background   sources from the list of \citet{boogert}. 
The light from the brightest  stars contaminate large areas that  extends into the annulus where $CS$ is measured. 
We  therefore excluded sources brighter than eight magnitude at 3.4 $\mu$m obtained from the  WISE point source catalogue. 
The limit was chosen by finding no contamination under visual inspection, and also by testing the measured $CS$ surface brightness for variations in the annulus. 
The final sample contains 12 sources. 
This selection is  not expected to bias the results since the objects are background objects and there is no preferred location within the cores. 
Although lower column densities preferentially were removed, we preserved essentially the whole extinction range from the  sample ($0.12 <\tau_{9.7} < 0.55$ for the reduced sample versus  $0.08 <\tau_{9.7} < 0.55$ for the full sample). 
As shown below, the correlations between water-ice and dust are similar for the sub-sample as for the full sample.

To measure the absolute $CS$ surface brightness, the relative zero point of the frames has to be known. 
We used the regions in the map without any $^{13}\mathrm{CO}(2-1)$ emission detected in Fig.~\ref{overview} to estimate  the background. 
The background values in the northern, western, and eastern parts all agreed within 0.005 MJy/sr, suggesting the background is flat, and we  used a single value for the whole map. 

\section{Results and discussion} 
We present correlations between the optical depths of the 3.0 $\mu$m ice feature, the 9.7 $\mu$m silicate feature, and the measured $CS$.

\subsection{Corrections for optical depth effects} 
The measured $CS$ surface brightness depends on the optical depth of the line of sight, and some self-absorption is expected. 
This has to be corrected to obtain the amount of $CS$ per unit dust. 
The correction is  a complicated interplay of the relative geometry of the sight line through the cloud and the structure of the cloud. 

To quantify the correction as a function of optical depth and location within the cloud, we  calculated a set of model cores assuming they are located at the Galactic position of the Lupus IV complex. 
The model core parameters were the same as in \citet{andersen} except that the mass was varied to emulate different optical depths. 
The line-of-sight radiation field was determined to be 0.17$\pm$0.04 MJy/sr from the combination of the DIRBE map and the WISE catalogue, as detailed in \citet{andersen}. 
Model cores were chosen to cover a range of maximum peak column densities, with the highest value being A$_{3.6}=3.4$, higher than the peak for most low-mass molecular cores. 
This will in turn contribute to an overestimate of the scatter because of too strong  shadowing for some sight-lines. 

The scattered-light contribution to the observed $CS$ depends on the orientation of the sight line relative to the interstellar radiation field. 
This is quantified by comparing the predicted $CS$ for all the different locations in the different models for a given optical depth. 
The average value was used for the correction and the uncertainty was included in the error estimates for each point.  
For an optical depth lower than unity the  standard deviation is around 30\%, but it increases to $\sim50$\% at higher optical depths.

\begin{figure}
 \resizebox{0.9\hsize}{!}{\includegraphics{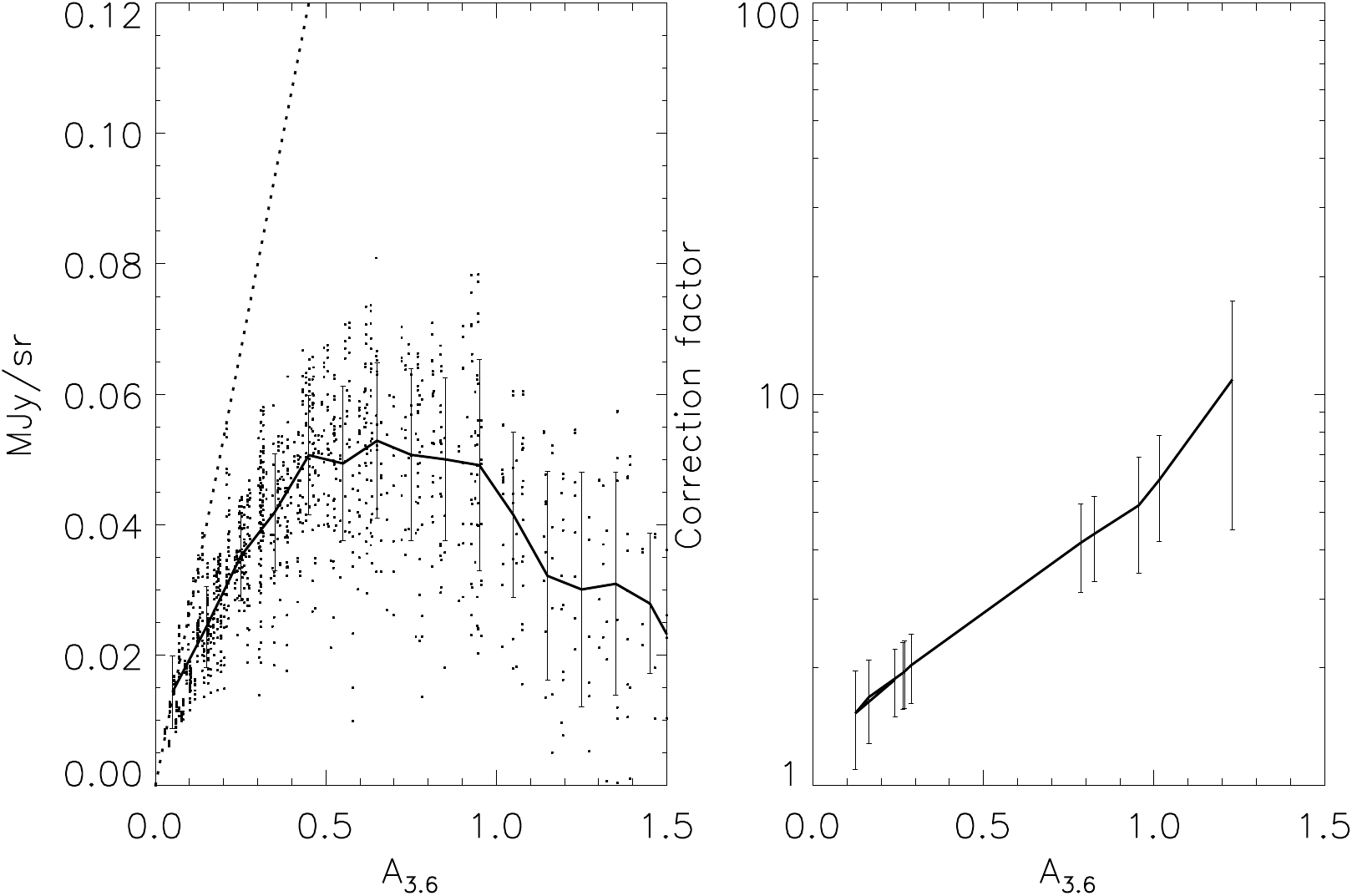}}
	\caption{Left: Surface brightness at 3.6 $\mu$m as a function of extinction for the model cores. The solid line is the average surface brightness. The dashed line is from a linear interpolation of the very optically thin case. Right: the correction factor for the adopted sample based on the ratio of the extrapolated and model surface brightness from the left. }
	\label{cor}
\end{figure}

Fig.~\ref{cor} shows the model surface brightness at 3.6 $\mu$m as a function of extinction  and  the predicted surface brightness extrapolated from the optical thin case (A$_{3.6}=0-0.07$). 
In the right panel in Fig.~\ref{cor} we show the correction (the ratio of the two lines in the left panel) for the extinction range of the sources in our sample. 
The \filter{K}- band extinction determined by  \citet{boogert}  was converted to 3.6 $\mu$m using their extinction laws. 
For A$_{K}< 1$, we adopted their low column density extinction law, and for A$_K>1$, their high column density law (in their Fig. 3).

\subsection{Coreshine versus ice and dust parameters}
We detect $CS$ emission above the background for all 12 sight lines, although for the lowest column density line-of-sight the measurement is marginal (0.0043 MJy/sr compared with a 1 $\sigma$ uncertainty in the background of 0.005 MJy/sr).  
Following the discussion in \citet{steinacker_new}, the location of Lupus IV at (l,b)=(336.7,+07.8)  indeed favours the scenario scattering exceeds the extinction of the background for a maximum grain size of 1 $\mu$m or more. 
The spectroscopic sample was not chosen specifically for sight lines where $CS$ was present above the background, and this  presents an unbiased selection of sight lines to measure $CS$.

\begin{figure}
 \resizebox{0.9\hsize}{!}{\includegraphics{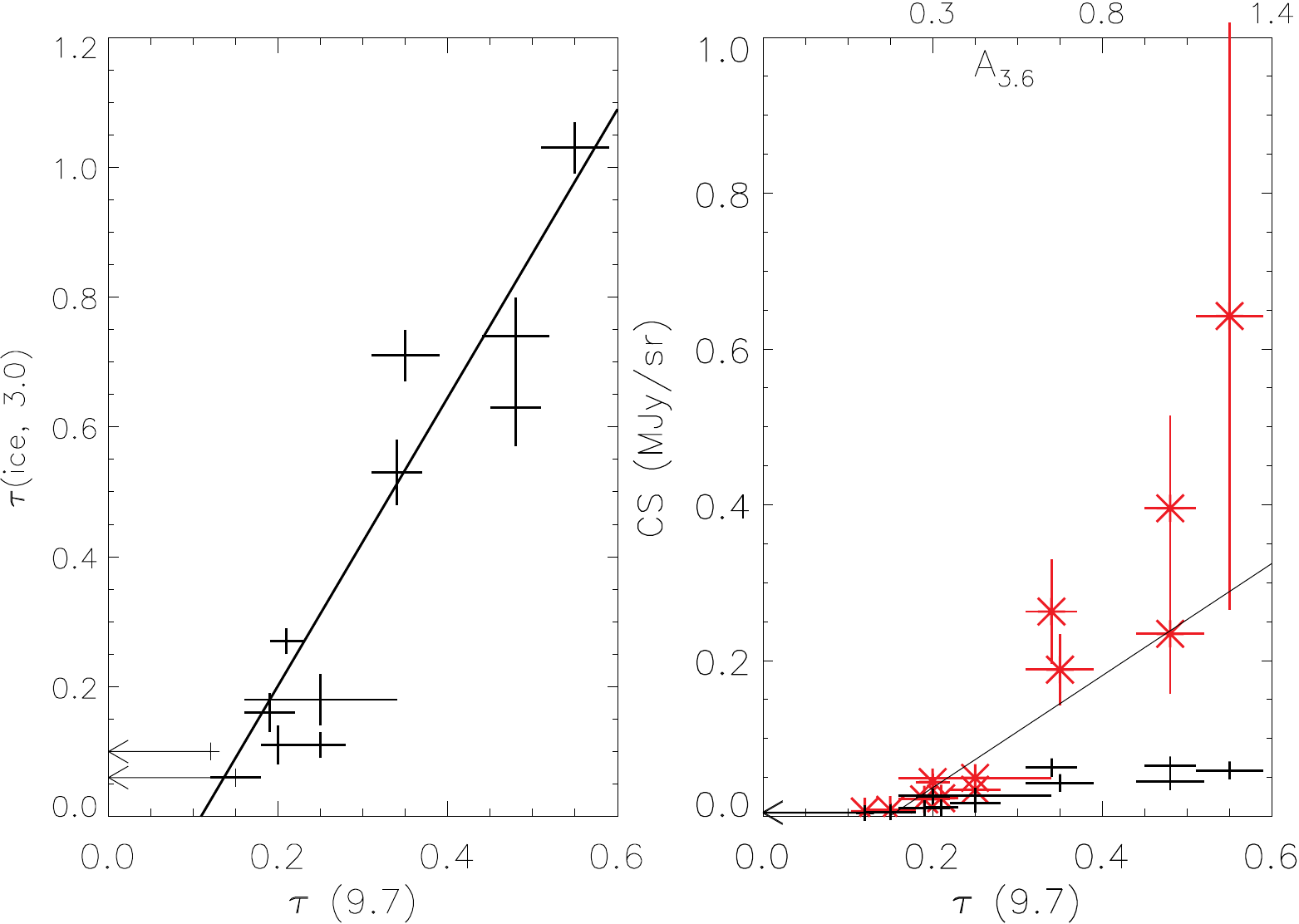}}
	\caption{Left:  optical depth of the ice feature as a function of the optical depth of the silicate feature. Right: measured $CS$ surface brightness and the corrected $CS$, $CS_{cor}$ as a function of the optical depth of the silicate feature. 
Plus signs with error bars are the measured values and red asterisk are the values corrected for optical-depth effects using the correction from Fig.~\ref{cor}.}
	\label{first}
\end{figure}

The left panel in Fig.~\ref{first} shows the optical depth at 9.7 $\mu$m versus  the optical depth of the water-ice feature at 3.0 $\mu$m for the sample. 
The increase in the water-ice optical depth with increasing  depth of the  silicate absorption feature is evident. 
The relationship between the $CS$ surface brightness and the silicate absorption shown  in the right panel in Fig.~\ref{first} follows a similar behaviour in that the $CS$ surface brightness increases when the optical depth at 9.7 $\mu$m increases.

\citet{boogert} presented the correlation between A$_{K}$ and $\tau^{ice}_{3.0}$. 
We  performed the same fit for the reduced sample here and find a similar  result, $\tau^{ice}_{3.0}=(-0.08\pm0.02)+(0.44\pm0.01)\cdot A_{K}$, as  for  the full sample ($\tau_{3.0}^{ice} = (-0.11\pm0.03)+(0.43\pm0.03)\cdot A_{K}$). 
We performed the same correlation for the extinction measured through the silicate feature. 
The same fit to the ice extinction using the silicate extinction as reference is $\tau^{ice}_{3.0}=(-0.24\pm0.02)+\tau_{9.7}\cdot (2.22\pm0.09)$. 
The subsequent correlations between ice and $CS$ were made both relative to the silicate optical depth and the extinction in the K band. 

Fig.~\ref{first} shows in the right panel the $CS$ as a function of the optical depth of the silicate feature. 
A similar fit to the corrected $CS$ surface brightness as a function of dust column density gives a best fit of $CS_{cor}=(-0.11\pm0.03)+(0.73\pm0.17)\cdot \tau_{9.7}$ MJy/sr. Using $A_K$ as reference the fit is $CS_{cor}=(-0.08\pm0.02)+(0.19\pm0.04)\cdot A_K$ MJy/sr. 
There appears to be an offset for $CS$ to be seen in emission over the background extinction similar to that observed for the ice feature. 
By extrapolating the fits to the intersection with the abscissa, we find a critical optical depth of $\tau_{9.7}=0.11\pm0.01, A_K=0.19\pm0.04$ for the onset of ice and $\tau_{9.7}=0.15\pm0.05, A_K=0.4\pm0.2$ for the onset of $CS$. 
The very  similar threshold for the two processes suggests a common origin. 
The threshold is furthermore fully compatible with that found for water-ice in LDN183 \citep{whittet_13}.

\subsection{Relative strength of coreshine}
\begin{figure}
 \resizebox{0.9\hsize}{!}{\includegraphics{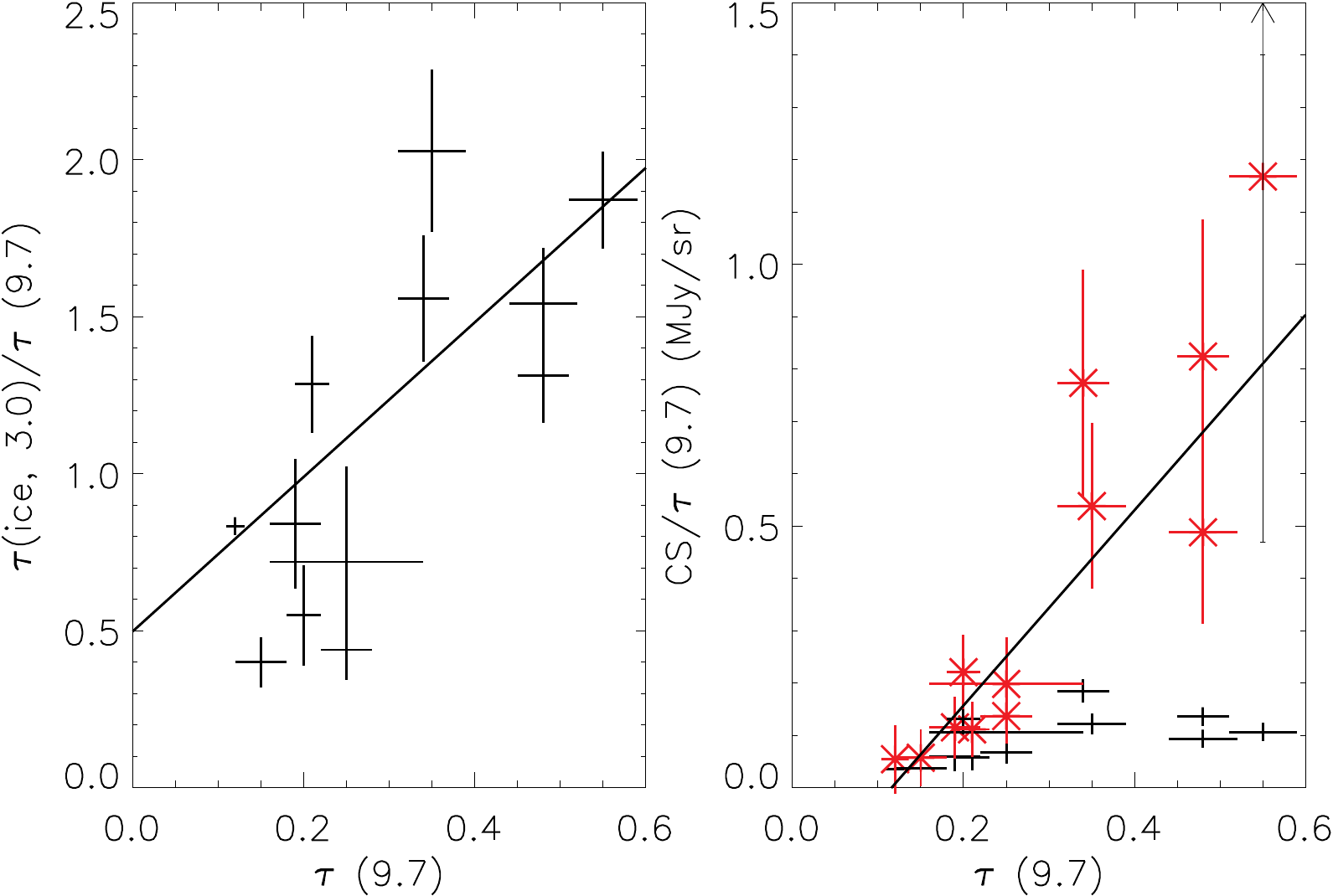}}	
	\caption{Left: relative optical depth from the ice feature as a function of $\tau_{9.7}$. Right:  relative $CS$ as a function of $\tau_{9.7}$. }
	\label{rel_dust}
\end{figure}

 Figure~\ref{rel_dust} indicates that  the ratio of ice abundance to dust increases with increasing column density. 
This was also seen for the full sample of sight lines in Lupus IV \citep{boogert}. 
However, it is unclear from the relation alone whether  this is caused by  a change in the dust properties or by a relative increase in the amount of water-ice. 
Assuming the latter, the corrected relative $CS$ surface brightness displays very similar behaviour, also rising strongly as a function of dust column density.
A linear fit to the relative water-ice abundance provides 
$\tau^{ice}_{3.0}/\tau_{9.7}=(0.5\pm0.18)+\tau_{9.7}\cdot (2.5\pm0.4)$, and for $A_K \tau^{ice}_{3.0}/A_K=(0.22\pm0.03)+A_K\cdot (0.08\pm0.02)$ and a similar fit to the corrected relative $CS$ surface brightness, $CS_{cor}$, provides  
$CS_{cor}/\tau_{9.7}=(-0.22\pm0.08)+\tau_{9.7}\cdot (1.9\pm0.4)$ using $\tau_{9.7}$ as a reference and $CS_{cor}/A_K=(-0.01\pm0.08)+A_K\cdot (2.0\pm0.17)$ for $A_K$.

\subsection{Grain growth and water-ice abundance}
\begin{figure}
 \resizebox{0.9\hsize}{!}{\includegraphics{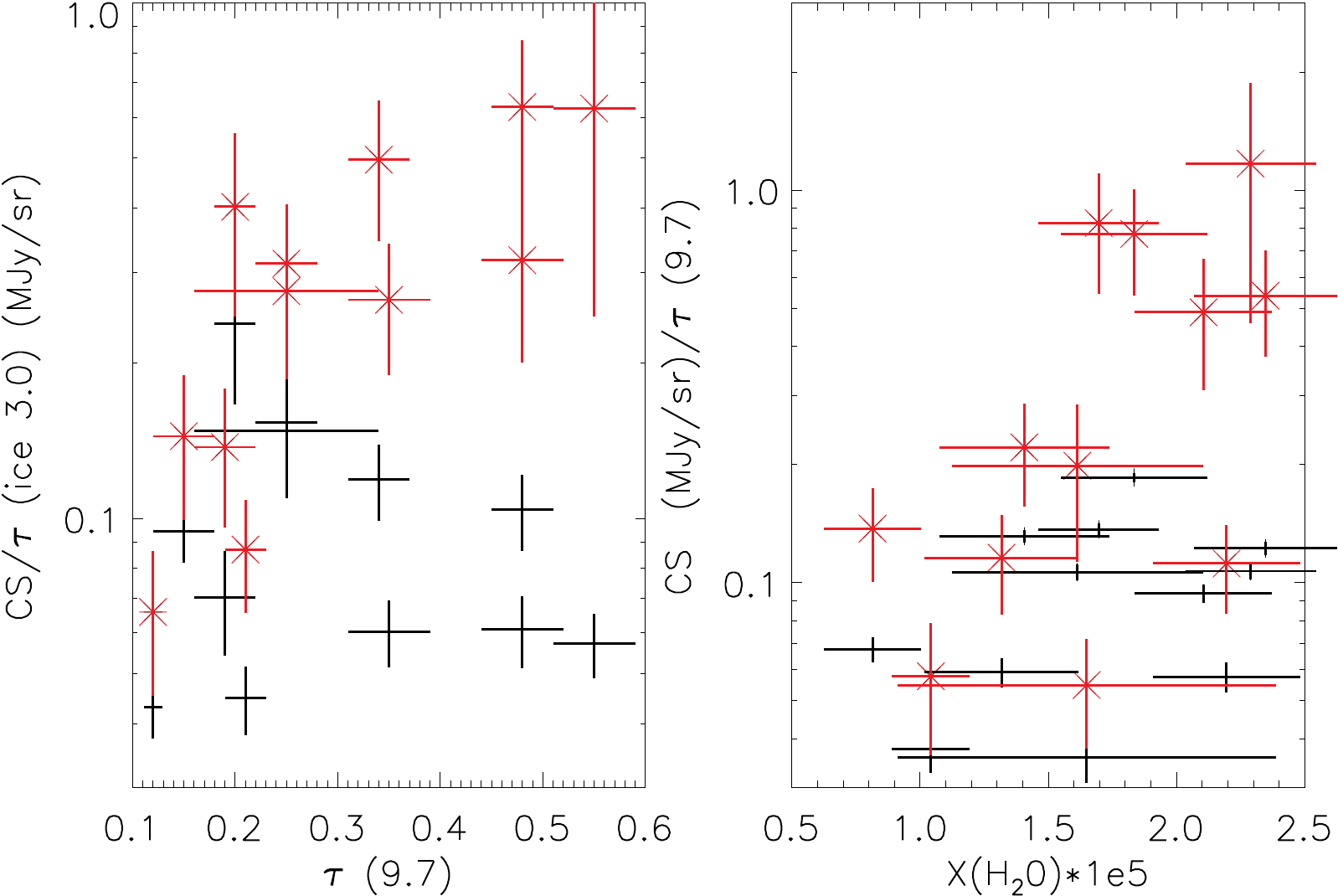}}
	\caption{Left:  ratio of $CS$ to $\tau^{ice}_{3.0}$ as a function of $\tau_{9.7}$. 
Right:  $CS$ surface brightness as a function of the relative ice water abundance.   }
	\label{water}
\end{figure}

We also  compared the ratio of the optical depth of the ice feature with the corrected $CS$ surface brightness in Fig.~\ref{water}. 
The ratio increases as a function of $\tau_{9.7}$ but there is substantial scatter. 
A clearer picture emerges from comparing the $CS_{cor}$ surface brightness with the relative amount of water-ice to hydrogen, shown in the right panel in  Fig.~\ref{water}. 
The  $CS_{cor}$ surface brightness increases with the relative water-ice abundance, $X(H_2O)$. 

Water-ice is commonly assumed to be the material necessary  for grain growth to be efficient through coagulation \citep[e.g.][]{ormel}. 
The grains grow through  layering of ice on their surface. 
Simple arguments suggest that the growth of the ice radius on each grain is  independent of the grain size, and thus the strongest effect would be for the small grains \citep{whittet} with a growth rate of $dr/dt\sim0.02\times(n_H/10^3)$ $\mu$m/Myr  \citep[  $n_H$ in $cm^{-3}$, e.g.][]{wick}. 
More complicated modelling supports this in the sense that the smallest grains grow relatively faster than large grains \citep{acharyya}. 
The effect would therefore be modest in terms of relative growth for micron-sized grains and is inefficient in itself to build the largest grains from 0.1 $\mu$m or smaller grains  that can be created through coagulation. 
The albedo can increase in any case for a power-law dust distribution with a \citet{MRN} slope. 
The scattering is proportional to $(2\pi a/\lambda)^4$, whereas the absorption only scales linearly in the Rayleigh-Taylor regime. Thus the albedo depends on the grain size roughly to the third power. For a power-law slope dust distribution the effect can roughly be estimated analytically. 
 The oxygen abundance in the ISM allows for an ice mantle thickness of up to 0.0175 $\mu$m \citep{draine85}.
For a grain size distribution in the range, for instance,  0.03-0.3 $\mu$m, the average grain size is 0.05 $\mu$m. Adding an ice mantle of  0.0175 $\mu$m,  the average grain size changes to 0.08 $\mu$m, resulting in a change in the albedo of 3.7.
More detailed Mie scattering calculations taking the full size distribution  and the optical properties of water-ice into account give an increase of a factor of 3.2.
Thus, the increased average size of  the grain size distribution is sufficient to increase the $CS$ efficiency. 
Simple ice covering will be efficient in increasing coreshine, but not in building  micron-sized grains. 
 Furthermore, thicker layers are possible if coagulation has already occurred or is occurring.

The correlations presented here suggest that  water-ice and $CS$ are related in molecular cores and that the $CS$ efficiency  may depend on the relative water-ice abundance. 
 The further increase of the $CS$ efficiency can either be caused by an increase of the ice-mantle thickness or coagulation. 
 Growth through both paths are  faster at higher densities.

If the water-ice is the pre-requisite for $CS$, one would expect different thresholds in regions with little to no water-ice. 
However, the stronger local radiation field has to be taken into account for the $CS$ in such an environment, which can be hidden in the relatively poor resolution of the current absolute measurements of the diffuse infrared radiation field.


\section{Conclusions}
We have combined measurements of the water-ice abundance along lines of sight through the Lupus IV region with measurements of scattering at 3.6 $\mu$m along the same lines of sight. 
For all the sight lines an excess emission over the background was found, which  was interpreted as scattering that is stronger than the extinction of the background radiation (coreshine). 
After correcting for optical depth effects, we found that the strength of $CS$ per unit dust increases with optical depth. 
A linear fit to both the optical depth of water-ice as a function of the optical depth of the silicate feature at 9.7 $\mu$m and the corrected $CS$ surface brightness revealed a minimum column density for  ice and $CS$. 
The thresholds for water absorption and $CS$ in emission are similar:  $\tau_{9.7}=0.11\pm0.01, A_K=0.19\pm0.04$ for water-ice and $\tau_{9.7}=0.15\pm0.14, A_K=-0.38\pm0.14$ for $CS$. 
The relative $CS$ strength increases with an increase of the relative water-ice abundance, suggesting that the $CS$ and ice feature are connected. 
The common increase of both relative $CS$ strength and water-ice absorption furthermore suggests a strong correlation between water-ice and $CS$. 
Studies of other molecular cloud complexes can reveal whether this is true in general and if there are environmental variations.

\begin{acknowledgements}
MA, WFT, and JS acknowledge support from the ANR (SEED ANR-11-CHEX-0007-01)
This publication makes use of data products from the Wide-field Infrared Survey Explorer, which is a joint project of the University of California, Los Angeles, and the Jet Propulsion Laboratory/California Institute of Technology, funded by the National Aeronautics and Space Administration. This work is based in part on observations made with the Spitzer Space Telescope, which is operated by the Jet Propulsion Laboratory, California Institute of Technology under a contract with NASA
\end{acknowledgements}

\begin{table}
\begin{tabular}{c c c c c }
\hline
Name & $CS (MJy/sr)$ & $\tau_{97}$ & $A_{K}$ & $\tau_{30}$\\
2MASS16014254-4153064 & 0.058 & 0.55$\pm$0.04 & 2.46$\pm$0.10 & 1.03$\pm$0.04\\
2MASS16004739-4203573 & 0.065 & 0.48$\pm$0.03 & 2.03$\pm$0.08 & 0.63$\pm$0.06\\
2MASS16010642-4202023 & 0.045 & 0.48$\pm$0.04 & 1.91$\pm$0.07 & 0.74$\pm$0.06\\
2MASS16012635-4150422 & 0.042 & 0.35$\pm$0.04 & 1.65$\pm$0.08 & 0.71$\pm$0.04\\
2MASS16012825-4153521 & 0.063 & 0.34$\pm$0.03 & 1.57$\pm$0.12 & 0.53$\pm$0.05\\
2MASS16021102-4158468 & 0.016 & 0.25$\pm$0.03 & 0.72$\pm$0.05 & 0.11$\pm$0.02\\
2MASS16014426-4159364 & 0.011 & 0.19$\pm$0.03 & 0.67$\pm$0.03 & 0.16$\pm$0.03\\
2MASS16022128-4158478 & 0.012 & 0.21$\pm$0.02 & 0.66$\pm$0.05 & 0.27$\pm$0.02\\
2MASS16000067-4204101 & 0.026 & 0.20$\pm$0.02 & 0.41$\pm$0.03 & 0.11$\pm$0.03\\
2MASS16005559-4159592 & 0.005 & 0.15$\pm$0.03 & 0.31$\pm$0.05 & 0.06$\pm$0.00\\
2MASS16003535-4209337 & 0.026 & 0.25$\pm$0.09 & 0.60$\pm$0.09 & 0.18$\pm$0.04\\
2MASS16024089-4203295 & 0.004 & 0.12$\pm$0.00 & 0.31$\pm$0.03 & 0.10$\pm$0.00\\
\hline
\hline
\end{tabular}
\caption{Tabulation of the $CS$ values and the corresponding optical depth measurements and extinction in the \filter{K} band from \citet{boogert}. The errors for the $CS$ measurements are dominated by the background uncertainty.}
\end{table}

\end{document}